\documentclass[lettersize,journal]{IEEEtran}
\usepackage{amsmath,amsfonts}
\usepackage{algorithmic}
\usepackage{algorithm}
\usepackage{array}
\usepackage[caption=false,font=normalsize,labelfont=sf,textfont=sf]{subfig}
\usepackage{textcomp}
\usepackage{stfloats}
\usepackage{url}
\usepackage{verbatim}
\usepackage{graphicx}
\usepackage{cite}
\usepackage{cite}
\usepackage{amsmath,amssymb,amsfonts}
\usepackage{algorithmic}
\usepackage{graphicx}
\usepackage{textcomp}
\usepackage{xcolor}
\usepackage{booktabs}

\usepackage{algorithmic}
\usepackage{algorithm}

\usepackage{listings}
\usepackage{xcolor} % Optional for custom colors

\def\BibTeX{{\rm B\kern-.05em{\sc i\kern-.025em b}\kern-.08em
		T\kern-.1667em\lower.7ex\hbox{E}\kern-.125emX}}
\begin{document}

\title{Neural Network Graph Similarity Computation Based on Graph Fusion}

%\author{IEEE Publication Technology,~\IEEEmembership{Staff,~IEEE,}
        % <-this % stops a space
\author{
	Zeng Hui Chang, Yiqiao Zhang and Hong Cai Chen \IEEEmembership{Member,~IEEE,} 

	School of Automation, Southeast Unviersity, Nanjing, China
	
	Email: chenhc@seu.edu.cn

\thanks{This paper was produced by the IEEE Publication Technology Group. They are in Piscataway, NJ.}% <-this % stops a space
\thanks{Manuscript received April 19, 2021; revised August 16, 2021.}}

\markboth{Journal of \LaTeX\ Class Files,~Vol.~14, No.~8, August~2021}%
{Shell \MakeLowercase{\textit{et al.}}: A Sample Article Using IEEEtran.cls for IEEE Journals}


\maketitle

\begin{abstract}
Graph similarity learning, crucial for tasks such as graph classification and similarity search, focuses on measuring the similarity between two graph-structured entities. The core challenge in this field is effectively managing the interactions between graphs. Traditional methods often entail separate, redundant computations for each graph pair, leading to unnecessary complexity. This paper revolutionizes the approach by introducing a parallel graph interaction method called graph fusion. By merging the node sequences of graph pairs into a single large graph, our method leverages a global attention mechanism to facilitate interaction computations and to harvest cross-graph insights. We further assess the similarity between graph pairs at two distinct levels—graph-level and node-level—introducing two innovative, yet straightforward, similarity computation algorithms. Extensive testing across five public datasets shows that our model not only outperforms leading baseline models in graph-to-graph classification and regression tasks but also sets a new benchmark for performance and efficiency. The code for this paper is open-source and available at https://github.com/LLiRarry/GFM-code.git

\end{abstract}


\section{Introduction}

Applications based on graph data structures, such as social network analysis\cite{1}, bioinformatics\cite{2}, and recommendation systems\cite{3}, play a crucial role in both academia and industry. Among them, graph similarity learning is one of the most significant challenges in graph-related machine learning tasks. These tasks include graph search in graph databases\cite{4}, malware detection\cite{5}, and brain data analysis\cite{6}. Traditional graph similarity computation methods such as Beam\cite{7}, Hungarian\cite{8}, VJ\cite{9}, A*\cite{10}, and HED\cite{11} use Graph Edit Distance (GED)\cite{12} or Maximum Common Subgraph (MCS)\cite{13} as similarity metrics and then employ exact or approximate techniques to calculate similarity scores. However, estimating GED or MCS between arbitrary graph structures is computationally expensive, a known NP-complete problem\cite{13,14}. Although approximate algorithms offer faster solutions, they may still be complex and operate in polynomial or sub-exponential time complexity, making them difficult to generalize for large-scale graphs in real-world applications. Moreover, the rich node feature information is often overlooked, thus losing potential semantic similarities between graphs.

In recent years, Graph Neural Networks (GNNs) have emerged as a highly promising approach for graph mining tasks, proving to be a powerful class of deep learning models. Models developed based on GNNs perform core operations of iteratively aggregating node representations from their structural neighborhoods, which can be applied to a variety of downstream applications, including but not limited to node classification \cite{15}, graph classification \cite{16}, and graph generation \cite{17}. In node classification, GNNs effectively learn node embedding vectors used directly for classification tasks; in graph classification, node embedding vectors are aggregated into graph-level embedding vectors; in graph generation, GNNs are combined with variational autoencoders to learn the distributional representations of graphs. Leveraging the strong representational capabilities of GNNs, GNN-based Graph Similarity Computation (GSC) solutions are becoming increasingly popular. To adapt GNNs for GSC tasks, target similarity scores, such as the GED, are normalized to the range (0, 1]. Thus, GSC can be viewed as a single-value regression problem, which outputs a similarity score by inputting two graphs.

Various Graph Neural Network architectures have explored mapping relationships between graph pairs and similarity scores \cite{18,19,20,21,22}. Techniques like histogram features \cite{18} and Convolutional Neural Networks for pairwise node similarity matrices \cite{20} have been used to compute node-level similarities. For graph-level embeddings, methods include Neural Tensor Networks \cite{18} and combining graph interactions with LSTM \cite{21} and Hypergraph Networks \cite{22} for efficient pooling. However, histogram features lack granularity, and focusing only on node or graph levels misses comprehensive structural similarities. Calculating graph interactions involves cross-level interactions between graph-to-node \cite{21}, node-to-node \cite{20}, and hyperedges in hypergraphs \cite{22}, each requiring extensive computation. Furthermore, interaction computations need to be reciprocated between graph pairs, doubling the computational effort. Enhancements like layering modules \cite{22} or using non-parallel recursive neural networks for pooling \cite{21} slow down inference speed.

Addressing the shortcomings of previous research, this study introduces a novel network for graph similarity calculation featuring a new graph fusion module for cross-graph information interaction. Graph fusion involves concatenating the node sequences of two graphs into a single large graph, which is then processed through a Transformer structure with a global attention mechanism for interactive encoding. After interaction, the large graph is split back into the original two graphs according to their original node sequences. This mechanism accounts for the relevance of nodes within and outside the graphs, achieving low-dimensional node-level interaction calculations while also merging global information of the graph pair to enhance the overall difference or similarity learning, thus enriching node information expression. Importantly, this graph interaction process enables simultaneous interactions within the graph pair, reducing computational load by half compared to previous efforts. A variant of Transformer, the Performer, which operates with linear complexity, is introduced to achieve linear complexity in graph interaction computations..

For downstream tasks, two new similarity learning networks are proposed: graph-level similarity learning using one-dimensional convolution and node-level similarity learning using one-dimensional grouped convolution. The learned similarity vectors from two perspectives are concatenated and input into a multilayer perceptron to obtain the final similarity score. The overall model includes node encoding, graph fusion, and two graph similarity learning modules, achieving optimal results with a straightforward structure.

In summary, the contributions are:

\textbf{(1)} Proposing a novel graph interaction method, graph fusion, with linear computational complexity, enabling simultaneous information interaction and expression optimization for two graphs. This method enhances node information learning by emphasizing the relevance or difference features of input graph pairs.

\textbf{(2)} Introducing two novel similarity learning modules: graph-level and node-level,  These multi-perspective computational methods enrich the model with richer information and stronger learning capabilities.

\textbf{(3)} Conducting comprehensive experiments on five datasets to validate the proposed model Graph Fusion Model(GFM) and using ablation studies to analyze various components. The model shows significant gains over state-of-the-art baselines.

\section{Related work}
\subsection{Graph Neural Networks}
Graph Neural Networks use deep neural network techniques to learn effective graph representations. They are divided into two main types: spectral methods and spatial methods. Spectral methods define graph convolution operations based on graph spectral theory. For example, \cite{25} uses the graph Laplacian's eigenvectors in the Fourier transform domain for graph convolution, \cite{26} approximates $K$-order polynomial spectral filters using Chebyshev polynomials, and GCN \cite{27} simplifies the Chebyshev polynomials to the first order for effective layer-by-layer propagation. In contrast, spatial methods define graph convolution by aggregating the spatial neighborhoods of nodes. For instance, GraphSAGE \cite{28} samples and aggregates representations from local neighborhoods in an inductive framework, while GAT \cite{29} introduces an attention mechanism to adaptively aggregate neighborhood representations of nodes.

\subsection{Graph Similarity Learning}
Graph similarity learning aims to find a function that measures the similarity between two graphs. Traditional methods like GED \cite{12} and MCS \cite{13} have exponential time complexity, limiting their use for large graphs. Graph kernel methods \cite{30} offer an alternative but require high computational and storage costs. Recently, Graph Neural Networks (GNNs) have been used for graph similarity learning. SimGNN \cite{18} uses histogram features and neural tensor networks \cite{31} to model interactions at node and graph levels, respectively. GraphSim \cite{19} extends SimGNN by incorporating convolutional neural networks to capture complex node-level interactions. GMN \cite{20} propagates node representations within each graph and across attention-based neighbors from another graph. HGMN \cite{21} compares node representations from one graph with the representation of the entire other graph to enable cross-graph interaction. H2MN introduces hypergraphs to model substructure similarity between graphs \cite{22}, and other works \cite{30, 31} segment graphs into subgroups for node-level comparisons.

\cite{47} proposed a learning-based method called Neural Supergraph Similarity Search (NSS) to address the hypergraph search problem. This method efficiently performs hypergraph search in the vector representation space by learning the representations of query and data graphs. In \cite{48}, a novel approach was introduced for efficient graph similarity search in large-scale graph databases, aiming to solve the graph similarity search problem under graph edit distance constraints. Similarly, \cite{49} also introduced a technique focused on efficient graph similarity search in large graph databases, specifically addressing the challenge of graph similarity under edit distance constraints. \cite{50} introduced C-MPGCN, a graph-based deep metric learning algorithm designed for regional similarity learning. This method aims to overcome the limitations of existing approaches, which often overlook spatial relationships and important features, by representing regions as graphs and using graph convolutional networks combined with contrastive learning to predict regional similarities. In \cite{51}, the authors proposed an algorithm named INFMCS (Similarity Computation via Maximum Common Subgraph Inference) for graph similarity computation. This algorithm seeks to address the lack of interpretability in existing learning methods for graph similarity measurement, by implicitly inferring the maximum common subgraph (MCS) to compute graph similarity, thereby making the process more interpretable.

\subsection{Graph Transformer}
The integration of Graph Neural Networks with Transformer architectures is increasingly being used to address challenges in graph networks. The existence of Graph Transformers has enhanced the performance of GNNs in handling long-range dependencies and large-scale graph data \cite{32,33}. The Simplified Graph Transformer (SGFormer) \cite{34} significantly improves performance and computational speed with just one layer of attention. The extensive, robust, and scalable GPS \cite{35} has expanded the model to various graph sizes and replaced the Transformer with a variant with linear computational complexity to enhance computational efficiency. Graph similarity calculation networks based on Graph Transformers \cite{36,37} have also been proposed, where Transformer modules are used to facilitate cross-graph information interaction, with Key and Value from one graph and Query from another, to perform node-level interaction calculations.

\section{The proposed model}

\subsection{Notations and Problem Formulation}
Consider a collection of graph pairs $\mathcal{G} = \{(\mathcal{G}_{1,i}, \mathcal{G}_{1,j}), (\mathcal{G}_{2,i}, \mathcal{G}_{2,j}), \dots, (\mathcal{G}_{n,i}, \mathcal{G}_{n,j})\}$, with each graph having an arbitrary number of nodes and edges. Each pair $(\mathcal{G}_{i}, \mathcal{G}_{j})$ includes a first graph $\mathcal{G}_{i} = (\mathcal{V}_{i}, \mathcal{E}_{i}, X_{i})$, where $n_{i} = |\mathcal{V}_{i}|$ represents the number of nodes and $e_{i} = |\mathcal{E}_{i}|$ represents the number of edges. An adjacency matrix $\mathcal{A}_{i} \in \mathbb{R}^{n_{i} \times n_{i}}$ describes the connections between nodes, while the node feature matrix $X_{i} \in \mathbb{R}^{n_{i} \times f}$ describes attributes of nodes, with $f$ being the attribute dimension. The second graph $\mathcal{G}_{j}$ is similarly defined as $\mathcal{G}_{j} = (\mathcal{V}_{j}, \mathcal{E}_{j}, X_{j})$. With these definitions in place, the problem setup is as follows:

\textbf{Input:} We are given a set of graph training pairs $\mathcal{G}_\mathcal{T}$ with corresponding supervision information $\mathcal{Y}_\mathcal{T}$

\textbf{Output:} The objective is to use a graph neural network to determine the similarities between arbitrary pairs of graphs in an end-to-end fashion, denoted by $y = \mathcal{S}(\mathcal{G}_{i}, \mathcal{G}_{j})$. This study focuses on two types of tasks based on graph-graph similarity: classification and regression. In the classification task, $y$ indicates a class label, specifically $y \in \{-1, 1\}$. In the regression task, $y$ quantifies the similarity between graphs, with $y \in (0, 1]$.
To well fit GNN, the standard GED value is normalized as
the $nGED$, i.e., 
${nGED}(\mathcal{G}_{i}, \mathcal{G}_{j}) = \frac{{GED}(\mathcal{G}_{i}, \mathcal{G}_{j})}{{(|\mathcal{G}_{i}| + |\mathcal{G}_{j}|)}\\/ {2}}.
$
In the following, $nGED$ should be transformed to the
value ranging $(0, 1]$ as the ground truth similarity score $s_{ij}$, i.e.,
$
s_{ij} = \exp(-{nGED}(\mathcal{G}_{i}, \mathcal{G}_{j})) \in S,
$
where $S \in \mathbb{R}^{|D| \times |D|}$ indicates the similarity matrix among all the graphs.

\subsection{The Overall Framework}
Figure 1 depicts our proposed model framework, which comprises four main parts: (1) Node Embedding Encoding for learning high-order node representations; (2) Graph Fusion for Cross-Graph Interaction; (3) Graph-level Similarity Calculation for learning the similarity of high-dimensional graph-level embedding vectors; (4) Node-level Similarity Calculation for learning the similarity of low-dimensional node-level embedding vectors. The model is structured into three sequential levels, with (3)(4) at the same level, and their outputs are concatenated and input into a multilayer perceptron to obtain the final similarity score or split label. Compared to existing methods, our model can capture rich substructural semantic similarities and is optimized end-to-end through backpropagation. Next, we will detail the principles of each part.

\begin{figure*}[t]
	\centering
	\includegraphics[width=0.8\textwidth]{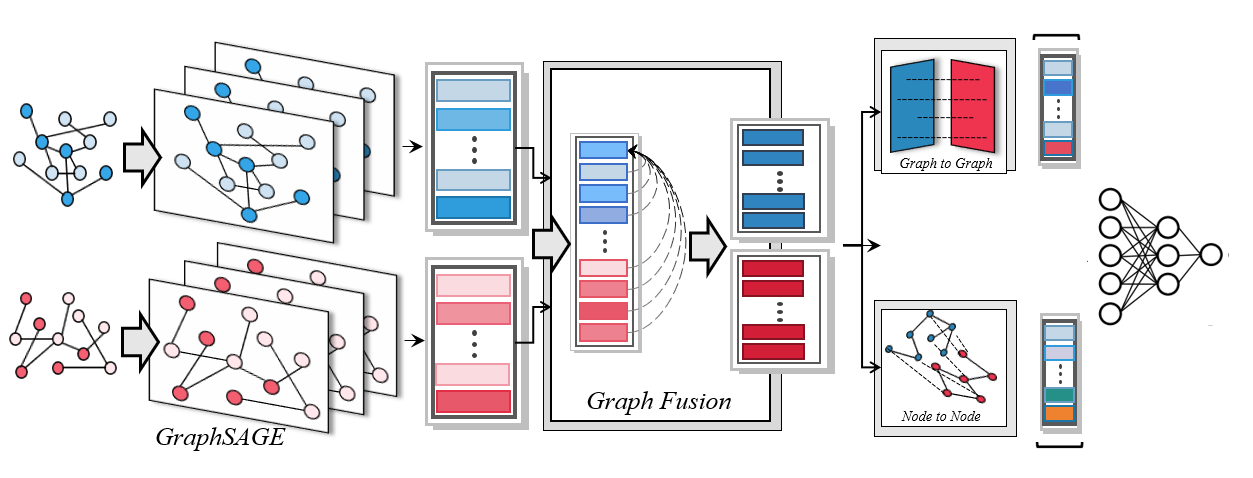}
	\caption{The overall architecture of our proposed GFM framework}
	\label{fig:screenshot003}
\end{figure*}

\begin{figure}
	\centering
	\includegraphics[width=0.99\linewidth]{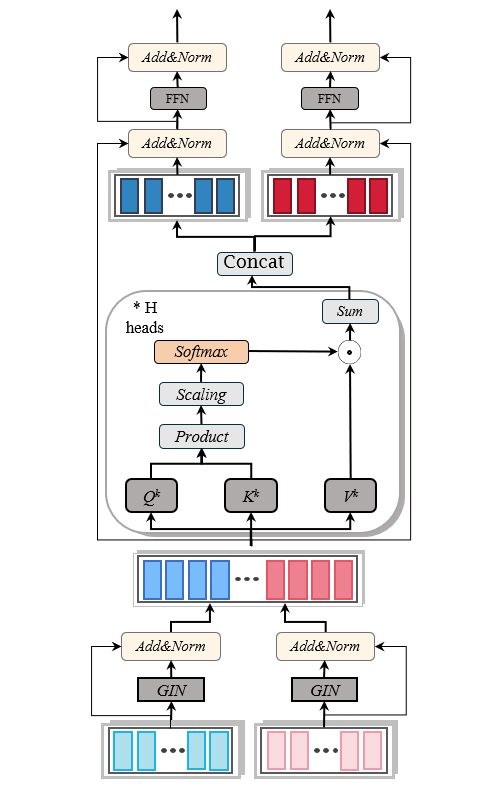}
	\caption{The overall architecture of our proposed Graph Fusion}
	\label{fig:screenshot002}
\end{figure}

\subsection{Node Embedding}

GraphSAGE \cite{38} is a framework for updating node information in graph neural networks. It uses node feature information to generate embeddings for unseen data by learning a function that samples and aggregates local neighborhood features. This approach has been effective for large graph datasets. The node update rules for GraphSAGE are shown as  (1) and (2).

\[\boldsymbol{h}_{\mathcal{N}(v)}^k = AGGREGAT{E_k}(\{ \boldsymbol{h}_u^{k - 1},\forall u \in \mathcal{N}(v)\} ) \tag{1}\]
\[\boldsymbol{h}_v^k = \sigma ({\boldsymbol{W}^k} \cdot CONCAT(\boldsymbol{h}_v^{k - 1},\boldsymbol{h}_{\mathcal{N}(v)}^k)) \tag{2}\]

The aggregation in \texttt{AGGREGATE} can use methods such as `\texttt{mean}`, `\texttt{max}`, or `\texttt{LSTM}`. This paper opts for `\texttt{mean}` aggregation as a balance between efficiency and effectiveness. It's important to obtain high-dimensional representations of graph nodes to integrate their structural and categorical information before engaging in deep interactive learning of low-dimensional graph structures. Therefore, three consecutive GraphSAGE layers are used for graph structure learning, similar to \cite{18}. However, continuous multi-layer node encoding may cause the loss of crucial node category information for similarity calculation, considering the importance of node category features and the presence of graph smoothing phenomenon \cite{40}.

\[\boldsymbol{h}_v^k = GraphSAG{E_{mean}}(\boldsymbol{h}_v^{k - 1}) + \boldsymbol{h}_v^{k - 1} \quad k \in \{ 2,3\} \tag{3} \]

Hence, inspired by the residual network concept from \cite{39}, residual connections are introduced in the node encoding layers to retain original features while deepening graph structure learning and to mitigate issues such as gradient vanishing, as shown in (3).

\subsection{Graph Fusion}
In previous work \cite{20}, a graph matching network was introduced to compute graph similarity. It modifies the node update module in each propagation layer to include a cross-graph matching vector, denoted as (4), which measures the degree of match between nodes in different graphs. Here, \(m_{j \rightarrow i}\) is the message vector from node \(j\) to node \(i\) within the same graph, and \(\sum_{j'} \mu_{j' \rightarrow i}\) is the sum of cross-graph matching vectors from all nodes \(j'\) in the other graph to node \(i\).

\[\boldsymbol{h}_i^k = GNN(\boldsymbol{h}_i^{k - 1},\sum\limits_{j \in \mathcal{N}(i)} {{m_{j \to i}},} \sum\limits_{j' \in {\mathcal{G}_{other}}} {{a_{j' \to i}}(\boldsymbol{h}_i^{k - 1}}  - \boldsymbol{h}_{j'}^{k - 1})) \tag{4}\]

However, this method requires interactive computations to be performed twice for each graph separately. It also lacks a mechanism for learning global features across graphs, as it only considers neighbor nodes within a graph while considering all nodes of the other graph, leading to an imbalance in information reception. Another approach \cite{21,22} uses nodes or hyperedges of one graph to represent nodes in the original graph across graphs. This process typically employs an attention mechanism for aggregation, followed by computing a one-to-one similarity vector between nodes of the original graph and the cross-graph aggregated nodes using cosine similarity, which is then aggregated into a graph-level similarity vector by a readout layer. However, this computation is also done twice since it requires interactive encoding for both graphs separately.

Notably, \cite{36,37} also proposed a Transformer-based graph interaction structure, which is different from the structure proposed in this paper. In the typical self-attention mechanism, Query ($Q$), Key ($K$), and Value ($V$) all originate from different projections of the same sequence, designed to evaluate the internal influences and dependencies among various parts of the sequence. In \cite{36,37}, ${Q_i} \in {\mathcal{G}_i}$,${K_j} \in {\mathcal{G}_j}$,${V_j} \in {\mathcal{G}_j}$ ,as shown in (5). $H$ represents the number of heads in the multi-head attention mechanism.

\[{\boldsymbol{\tilde {\mathcal{G}_i}}} = {\boldsymbol{\tilde Q_{i}}^k}\mathop {||}\limits_{k = 1}^H softmax (\frac{{\boldsymbol{Q}_{i}^k{{(\boldsymbol{K}_{j}^k)}^T}}}{{\sqrt {{d_k}} }})\boldsymbol{V}_{j}^k \tag{5}\]

This cross-graph attention mechanism achieves a result similar to the interactive encoding of \cite{21,22}, where the output is an expression of nodes in ${\mathcal{G}_i}$ that incorporates important information from nodes in ${\mathcal{G}_j}$. However, this graph interaction process requires interactive computations twice. Moreover, in \cite{36,37}, this optimized expression directly replaces the original encoding results of self-structured learning, which may lead to potential information loss.

The graph fusion module in this paper integrates cross-graph graph matching \cite{20} and interactive representation \cite{21,22}, allowing for interactive learning of structural and content information from the opposing graph within node embedding encoding. This process requires computation only once and simplifies the originally complex computational process to linear complexity by using the Performer instead of the Transformer. The computational flow diagram of this module is shown in Figure 2, and the pseudocode is presented as Algorithm 1. The input for the graph fusion part consists of the node sequences of two graphs, where $f$ represents the feature dimension and $|\mathcal{V}|$ represents the number of nodes. Initially, the nodes undergo enhanced independent learning through a layer of GIN \cite{41}, after which the node sequences from the two graphs are concatenated to construct a complete large graph. The cross-graph fusion interaction learning is achieved using a global attention mechanism encoding process, which can employ either a Transformer or a Performer. For data mining tasks, this linear approximation of interaction encoding significantly enhances model computational efficiency and reduces computational complexity.After the interactive fusion encoding, the large graph is split back into the original two input graphs according to their original node positions. The normalization method used in this study's graph fusion module is batch normalization.

\begin{algorithm}[H]
	\caption{Graph Fusion Model}
	\begin{algorithmic}[1]
		\STATE {\textbf{INPUT}}: \textit{Node features} ${\boldsymbol{X}_i} \in {\mathbb{R}^{\left| {{\mathcal{V}_{i }}} \right| \times f}}$ ,${\boldsymbol{X}_j} \in {\mathbb{R}^{\left| {{\mathcal{V}_{j}}} \right| \times f}}$
		\STATE {\textbf{OUTPUT}}: \textit{Updated node features after fusion} ${\boldsymbol{ \tilde X_i}} \in {\mathbb{R}^{\left| {{\mathcal{V}_{i }}} \right| \times f}}$ ,${\boldsymbol{ \tilde X_j}} \in {\mathbb{R}^{\left| {{\mathcal{V}_{j}}} \right| \times f}}$
		\STATE \textit{Independent encoding using GIN}

		$GIN: \boldsymbol{h}_v^k = MLP((1 + {\varepsilon ^{(k)}}) \cdot \boldsymbol{h}_v^{k - 1} + \sum\limits_{j \in \mathcal{N}(i)} {\boldsymbol{h}_j^{k - 1}))} $
		
		${\boldsymbol{\hat X_{i,j}}} = Norm(GIN({\boldsymbol{X_{i,j}}}) + {\boldsymbol{X_{i,j}}})$
		\STATE \textit{Merge node sequences from graphs} $\mathcal{G}_{i}$   \textit{and} $\mathcal{G}_{j}$  \textit {to form one graph}
		
		${\boldsymbol{\bar X}} = CONCAT({ \boldsymbol{\hat X_i}},{\boldsymbol{ \hat X_j}}) \in {\mathbb{R}^{\left| {({\mathcal{V}_i} + {\mathcal{V}_j}) \times f} \right|}}$
		\STATE $ \boldsymbol{Q} \leftarrow {\boldsymbol{\bar X}},\boldsymbol{K} \leftarrow {\boldsymbol{\bar X}},\boldsymbol{V} \leftarrow {\boldsymbol{\bar X}}$
		\STATE  \textit{If using Transformer for global attention mechanism}
		
		$\boldsymbol{\tilde{ \bar X}} = softmax (\frac{{\boldsymbol{Q}{\boldsymbol{K}^T}}}{{\sqrt {{d_k}} }})\boldsymbol{V}$
		\STATE  \textit{If using Performer for global attention mechanism}
		
		$\boldsymbol{\tilde{ \bar X}} = {\boldsymbol{\hat D}^{ - 1}}(\boldsymbol{Q}'({(\boldsymbol{K}')^T}\boldsymbol{V})),  \boldsymbol{\hat D} = diag(\boldsymbol{Q}'({(\boldsymbol{K}')^T}{1_L}))$
		\STATE \textit{Split and restore into the original two graphs} $\mathcal{G}_{i}$  \textit{and} $\mathcal{G}_{j}$
		
		${\boldsymbol{\tilde X_i}},{\boldsymbol{\tilde X_j}} = Split(\boldsymbol{\tilde {\bar X}})$,  ${\boldsymbol{\tilde X_i}} \in {\mathbb{R}^{\left| {{\mathcal{V}_{i }}} \right| \times f}}$ ,${\boldsymbol{\tilde X_j}} \in {\mathbb{R}^{\left| {{\mathcal{V}_{j }}} \right| \times f}}$ 
		
		\STATE ${\boldsymbol{\tilde X_{i,j}}} = Norm({\boldsymbol{\tilde X_{i,j}}} + {\boldsymbol{\hat X_{i,j}}})$
		\STATE 
		$ {\boldsymbol{\tilde X_{i,j}}} = Norm(ReLU(W({\boldsymbol{\tilde X_{i,j}}}) + {\boldsymbol{\tilde X_{i,j}}}))$
		
		\STATE \textbf{RETURN:} ${\boldsymbol{ \tilde X}_i \in {\mathbb{R}^{\left| {{\mathcal{V}_{i }}} \right| \times f}},\boldsymbol{\tilde X}_j} \in {\mathbb{R}^{\left| {{\mathcal{V}_{j }}} \right| \times f}}$
	\end{algorithmic}
\end{algorithm}

\subsection{Different Levels of Similarity Calculation}

\subsubsection{Graph-level Similarity}
Pooling layers in many studies capture graph-level encodings, which are then processed through modules like Neural Tensor Network (TNT) \cite{18} and Embedding Fusion Network (EFN) \cite{42} to approximate graph-level embedding similarity. Unlike direct similarity calculations, these methods aim to integrate and transform paired graph-level vectors into new encodings. The variance in these outputs can be substantial due to the similarity or difference in input graph pairs, or disparities between the test and training sets, potentially affecting model stability. We introduce a novel method to compute graph-level encoding similarity that yields stable and meaningful similarity measurements.

\[{\boldsymbol{h}_i} = \sum\limits_{n = 1}^{\left| {{\mathcal{V}_i}} \right|} {\sigma (\boldsymbol{x}_n^T\tanh ((\frac{1}{{\left| {{\mathcal{V}_i}} \right|}}} \sum\limits_{m = 1}^{\left| {{\mathcal{V}_i}} \right|} {{\boldsymbol{x}_m}) \boldsymbol{W})){\boldsymbol{x}_n}} \tag{6}\]

Initially, we use a \texttt{readout} method to obtain graph-level embedding vectors from node embedding vectors, as shown in (6), where \(W\) represents learnable weight parameters and the activation function used is \texttt{sigmoid()}. We employ two methods to compute the similarity scores of the embedded vectors: kernel similarity (7)(8) and an approximate form of hamming similarity (9).

%\[\boldsymbol{si{m_{(1)}}} = \frac{{\sum\limits_{k = 1}^{k = H} {{\boldsymbol{h}^{k}_i} \cdot {\boldsymbol{h}^{k}_j}} }}{{\sqrt {\sum\limits_{k = 1}^H {{{({\boldsymbol{h}^{k}_i})}^2}} } \sqrt {\sum\limits_{k = 1}^H {{{({\boldsymbol{h}^{k}_j})}^2}} } }} \tag{7}\]

\[\boldsymbol{d(h_i^k,h_j^k)} = \sqrt {{{({\boldsymbol{h}^{k}_i} - {\boldsymbol{h}^{k}_j})}^T}({\boldsymbol{h}^{k}_i} - {\boldsymbol{h}^{k}_j})} \tag{7} \]

\[\boldsymbol{si{m_{(1)}}} = {\boldsymbol{d(h_i^k,h_j^k)} \over {{\sigma ^2}}} \tag{8} \]

\[\boldsymbol{si{m_{(2)}}} = \frac{1}{H}\sum\limits_{k = 1}^H {\tanh ({\boldsymbol{h}^{k}_i}) \cdot } \tanh ({\boldsymbol{h}^{k}_j}) \tag{9}\]

However, similarity scores computed from a single high-dimensional perspective can be arbitrary and unstable \cite{43}, so this paper uses multiple sets of different one-dimensional convolution kernels and strides to project the original high-dimensional space of the graph embedding encoding into different low-dimensional spaces and computes multiple low-dimensional similarity scores. The resulting similarity scores are then concatenated to form a similarity score vector that reflects similarity from multiple perspectives. For a given one-dimensional input sequence \(x\), with length \(L\), convolution kernel \(w\), with length \(M\), stride \(S\), and \(H\) output channel. \(H\) can be considered as a method involving multi-head computation, where the mean of the similarity scores of the convolution results for \(H\) channels is taken as the output. The one-dimensional convolution operation can be represented by (10)(11).

%\[\boldsymbol{\tilde h[n]} = \sum\limits_{m = 0}^{M - 1} {\boldsymbol{\omega} [m] \cdot \boldsymbol{h}[n \cdot S + m]}\tag{9} \]
%\[\boldsymbol{\tilde h} = [\boldsymbol{\tilde h}[0],\boldsymbol{\tilde h}[1],...,\boldsymbol{\tilde h}[\frac{{L - M}}{S}]] \tag{10}\]

\[
\boldsymbol{\tilde h[n]} = \sum\limits_{m = 0}^{M - 1} w[m] \cdot x[n \cdot S + m] \tag{10}
\]

\[
\boldsymbol{\tilde h} = [\boldsymbol{\tilde h}[0], \boldsymbol{\tilde h}[1], \ldots, \boldsymbol{\tilde h}[\left\lfloor\frac{L - M}{S}\right\rfloor]] \tag{11}
\]

In this paper, the length of the graph-level embedding vector is 48, with convolution kernel lengths of \texttt{[1,2,3,4,5,6,7,8]} and strides of \texttt{[1,2,3]}. Each set of convolution kernels and strides corresponds to two similarity scores. This module, which projects the original high-dimensional space into multiple low-dimensional spaces from different angles and calculates multiple perspective similarity scores, outputs a real physical sense of similarity.

\subsubsection{Node-level Similarity}
Relying solely on similarity scores of graph-level embeddings is not thorough, as pooling algorithms often lose critical information when encoding from nodes to graphs. Previous research has explored low-dimensional node-level similarity calculations to gauge overall graph-level similarity, utilizing histogram features \cite{18} or learning correlation matrices with convolutional neural networks for node pairs \cite{19}. However, histogram features lack differentiability and fail to capture detailed features, while correlation matrices necessitate comparisons across all node pairs in two graphs, complicating computations due to the inherent complexity of graph data.

In this paper, we propose for the first time an algorithm based on grouped convolution to compute node-level similarity, assuming that the graph fusion module outputs the node sequences of two graphs as 
${\boldsymbol{X}_i} \in {\mathbb{R}^{B \times L \times f}}$ and ${\boldsymbol{X}_j} \in {\mathbb{R}^{B \times L \times f}}$, where $B$ represents the batch size, $L$ is $\max (\left| {\mathcal{V}_i} \right|,\left| {\mathcal{V}_j} \right|)$, and $f$ the feature dimension of the nodes. This part requires aligning the nodes of the two graphs using the \texttt{padding} method. The computation process of grouped convolution is as follows:

%\quad \quad  \texttt{\textbf{1. X=tanh(reshape(X,(1,B×f,L)))}}
%
%\quad \quad  \texttt{\textbf{2. kernels=tanh(reshape(Y,(B×f,1,L)))}}
%
%\quad \quad  \texttt{\textbf{3. result=Conv1d(X,kernels,groups=B×f)}}
%
%\quad \quad  \texttt{\textbf{4. output=reshape(result,(B,f))}}
%

\begin{lstlisting}
X = tanh(reshape(X, (1, B*f, L)))
kernels = tanh(reshape(Y, (B*f, 1, L)))
result = Conv1d(X, kernels, groups=B*f)
output = reshape(result, (B, f))
\end{lstlisting}

The \texttt{output} is a similarity vector at the node level with dimensions $f$, where \texttt{X} and \texttt{kernels} come from the node sequences of the two input graphs. The similarity of features across different dimensions is computed independently through grouped convolution.

To enhance the model's expressiveness and adaptability to this module, we incorporate batch normalization directly to the node sequences output by the graph fusion module. Batch normalization is used to stabilize the learning process by normalizing the outputs, reducing internal covariate shift. This approach improves training speed and helps the model generalize better by ensuring consistent distribution of input features across different training batches. The \texttt{tanh} activation function is used to capture similarity features in the computations of this module, rather than mere numerical integration.

\subsection{The  Final Concatenation and Output Layer}

We have computed and analyzed the graph similarity relationships of input graph pairs from three different perspectives. Next, the graph similarity vectors obtained from these perspectives will be concatenated. The concatenated vector is then fed into a multilayer perceptron (MLP) with a \texttt{sigmoid} activation function, to perform graph-graph classification and graph-graph regression tasks. This process and the loss function are defined as follows.

\[\boldsymbol{Z} = CONCAT(\boldsymbol{Sim(1)},\boldsymbol{Sim(2)}) \tag{12}  \]
\[\boldsymbol{\hat Y} = sigmoid(MLP(\boldsymbol{Z}))  \tag{13} \]
\[{\mathcal{L}_c} =  - \frac{1}{N}\sum\limits_{i = 1}^N {[{\boldsymbol{y}_i}\log ({{\boldsymbol{\hat y}}_i}) + (1 - {\boldsymbol{y}_i})\log (1 - {{\boldsymbol{\hat y}}_i})} ]  \tag{14}  \]
\[{\mathcal{L}_r} = \frac{1}{N}\sum\limits_{i = 1}^N {{{({\boldsymbol{y}_i} - {{\boldsymbol{\hat y}}_i})}^2}}  \tag{15} \]

where \(\mathcal{L}_c\) represents the binary cross-entropy loss for the graph-graph classification task, and \(\mathcal{L}_r\) is the mean square error loss for the graph-graph regression task. \(\boldsymbol{y}\) denotes the ground-truth supervision information.

\section{Experiments and analysis}

\subsection{Dataset}

We assessed our model using five public datasets, summarized in Table 1. These datasets, derived from FFmpeg and OpenSSL \cite{45,46}, represent binary function control flow graphs. We compiled source code functions with various compilers (e.g., gcc, clang) and optimization levels to create multiple binary function graphs. Binary functions from the same source are labeled semantically similar $({\mathcal{S}}({\mathcal{G}_i}, {\mathcal{G}_j}) = +1)$ and from different sources as dissimilar $({\mathcal{S}}({\mathcal{G}_i}, {\mathcal{G}_j}) = -1)$, framing the task as a graph-graph classification. Each dataset was divided into three subsets to analyze the impact of graph size.

\begin{table}[h]
	\caption{Statistics of the datasets. 3, 20, and 50 represent the minimum number of nodes in each subset. B (Billion), M (Million), K (Thousand).}
	\centering
	\small % Making the text a bit smaller
	\setlength\tabcolsep{3pt} % Default value: 6pt
	\begin{tabular}{@{}l|c|c|c|c|r@{}} % six columns
		\toprule
		\textbf{Datasets} & \textbf{Subsets} & \textbf{|$\mathcal{G}$|} & \textbf{Avg.|$\mathcal{V}$|} & \textbf{Avg.|$\mathcal{E}$|} & \textbf{ Graph Pairs} \\
		\midrule
		\textbf{FFmpeg}  & $[3, 200]$ & 83,008 & 18.83 & 27.02 & 6.89B \\
		& $[20, 200]$ & 31,696 & 51.02 & 75.88 & 1B \\
		& $[50, 200]$ & 10,824 & 90.93 & 136.83 & 117M \\
		\midrule
		\textbf{OpenSSL} & $[3, 200]$ & 73,953 & 15.73 & 21.97 & 5.46B \\
		& $[20, 200]$ & 15,800 & 44.89 & 67.15 & 249M \\
		& $[50, 200]$ & 4,308 & 83.68 & 127.75 & 18.5M \\
		\midrule
		\textbf{AIDS}   & -          & 700    & 8.90   & 8.80   & 490K \\
		\textbf{LINUX}  & -          & 1,000  & 7.58   & 6.94   & 1M \\
		\textbf{IMDB}   & -          & 1,500  & 13.00  & 65.94  & 2.25M \\
		\bottomrule
	\end{tabular}
\end{table}

In our research, we utilize the AIDS, LINUX, and IMDB datasets \cite{18} for graph-graph regression tasks, representing chemical compounds, program functions, and ego-networks, respectively. Each dataset includes actual Graph Edit Distance (GED) scores between graph pairs. For AIDS and LINUX, we compute GED precisely using the A* algorithm \cite{18}. For the IMDB dataset, with its higher computational demands, we employ approximate methods such as HED \cite{11}, Hungarian \cite{8}, and VJ \cite{9}, taking the lowest values as actual GED scores. Additionally, we normalize GED into a 0-1 similarity scale as outlined in Section 3.1.

\begin{table*}[t]
	\centering
	\caption{Graph-Graph classification results in terms of AUC score with standard deviation (in percentage).}
	\begin{tabular}{lcccccccccc}
		\toprule
		\midrule
		& \multicolumn{3}{c}{\textbf{FFmpeg}} & \multicolumn{3}{c}{\textbf{OpenSSL}} \\
		\cmidrule(lr){2-4} \cmidrule(lr){5-7}
		\textbf{Models} & [3, 200] & [20, 200] & [50, 200] & [3, 200] & [20, 200] & [50, 200] \\
		\midrule
		\midrule
		SimGNN [18] & 95.38±0.76 & 94.32±1.01 & 93.45±0.54 & 95.96±0.31 & 93.38±0.82 & 94.25±0.85 \\
		GMN [20] & 94.15±0.62 & 95.92±1.38 & 94.76±0.45 & 96.43±0.61 & 93.03±3.81 & 93.91±1.65 \\
		GraphSim [19] & 97.46±0.30 & 96.49±0.28 & 94.48±0.73 & 96.84±0.54 & 94.97±0.98 & 93.66±1.84 \\
		HGMN [21] & 98.07±0.06 & 98.29±0.10 & 97.83±0.11 & 96.90±0.10 & 97.31±1.07 & 95.87±0.88 \\
		H2MN(RW)[22] & 98.25±0.11 & 98.54±0.14 & 98.30±0.29 & 97.86±0.39 & 98.47±0.38 & 96.80±0.95 \\
		H2MN(NE)[22] & 98.28±0.20 & 98.31±0.14 & 98.16±0.56 & 98.27±0.16 & 98.05±0.36 & 97.78±0.75 \\
		\midrule
		\midrule
		\textbf{GFM(Transfromer)} & 98.62±0.11 & \textbf{99.03±0.24} & \textbf{99.21±0.10} & \textbf{99.18±0.14} & \textbf{99.22±0.15} & \textbf{98.99±0.55} \\
		\textbf{GFM(Performer)} & \textbf{99.03±0.09} & 98.89±0.10 & 98.74±0.29 & 99.08±0.16 & \textbf98.44±0.16 & 98.21±0.15 \\
		\midrule
		\bottomrule
	\end{tabular}
\end{table*}

\begin{table*}[t]
	\caption{Graph-Graph regression results for graph-to-graph regression in terms of mse($\times 10^{-3}$), $\rho$ and p@10. We use bold to highlight wins.}
	\centering
	\begin{tabular}{@{}lccccccccccc@{}}
		\toprule
		\midrule
		& \multicolumn{3}{c}{\textbf{AIDS}} & \multicolumn{3}{c}{\textbf{LINUX}} & \multicolumn{3}{c}{\textbf{IMDB}} \\
		\cmidrule(lr){2-4} \cmidrule(lr){5-7} \cmidrule(lr){8-10}
		\textbf{Models} & mse($\times 10^{-3}$) & $\rho$ & p@10 & mse($\times 10^{-3}$) & $\rho$ & p@10 & mse($\times 10^{-3}$) & $\rho$ & p@10 \\
		\midrule
		\midrule
		
		A* [10] & 0.000* & 1.000* & 1.000* & 0.000* & 1.000* & 1.000* & - & - & - \\
		Beam [7] & 12.090 & 0.609 & 0.481 & 9.268 & 0.827 & 0.973 & 2.413 & 0.905 & 0.813 \\
		Hungarian [8] & 25.296 & 0.510 & 0.360 & 29.805 & 0.638 & 0.913 & 1.845 & 0.932 & 0.825 \\
		VJ [9] & 29.157 & 0.517 & 0.310 & 63.863 & 0.581 & 0.287 & 1.831 & 0.934 & 0.815 \\
		HED [11] & 28.925 & 0.621 & 0.386 & 19.553 & 0.897 & 0.982 & 19.400 & 0.751 & 0.801 \\
		\midrule
		\midrule
		SimGNN [18] & 1.376 & 0.824 & 0.400 & 2.479 & 0.912 & 0.635 & 1.264 & 0.878 & 0.759 \\
		GMN [20] & 4.610 & 0.672 & 0.200 & 2.571 & 0.906 & 0.888 & 4.422 & 0.725 & 0.604 \\
		GraphSim [19] & 1.919 & 0.849 & 0.446 & 0.471 & 0.976 & 0.956 & 0.743 & \textbf{0.926} & 0.828 \\
		HGMN [21] & 1.169 & 0.905 & 0.456 & 0.439 & 0.985 & 0.955 & 0.335 & 0.919 & 0.837 \\
		H2MN(RW) [22]& 0.957 & 0.877 & 0.517 & 0.227 & 0.984 & 0.953 & \textbf{0.308} & 0.912 & 0.861 \\
		H2MN(NE)[22] & \textbf{0.913} & 0.881 & 0.521 & 0.105 & 0.990 & 0.975 & 0.589 & 0.913 & \textbf{0.889} \\
		\midrule
		\midrule
		\textbf{GFM(Transformer)} & 1.114 & \textbf{0.915} & \textbf{0.671} & \textbf{0.058} & \textbf{0.997} & \textbf{0.999} & 0.321 & 0.917 & 0.879 \\
		\textbf{GFM(Performer)} & 1.136 & 0.875 & 0.545 & 0.102 & 0.990 & 0.979 & 0.342 & 0.919 & 0.888 \\
		\midrule
		\bottomrule
	\end{tabular}
\end{table*}

\subsection{Baselines and Experimental Settings}

\noindent \textbf{\textit{Classical methods.}} These include algorithms for calculating the Graph Edit Distance (GED) like A* \cite{10}, Beam \cite{7}, Hungarian \cite{8}, VJ \cite{9}, and HED \cite{11}.

\noindent \textbf{\textit{Graph Neural Networks.}} This category comprises benchmarks such as SimGNN \cite{18}, GMN \cite{20}, GraphSim \cite{19}, HGMN \cite{30}, and H2MN \cite{22}, tested in classification and regression tasks.

\noindent \textbf{\textit{GFM variants.}} We consider GFMT using Transformer self-attention, and GFMP using the Performer's attention with linear complexity for graph fusion.

\noindent \textbf{\textit{Experiments and Parameter Settings:}} Datasets were divided into training, validation, and test sets with ratios of 80\%, 10\%, and 10\% for classification, and 60\%, 20\%, and 20\% for regression, following \cite{45,18}. Model parameters included a dimension of 48, 4 attention heads, and experiments were conducted on a platform with an Intel(R) Core(TM) i9-12900K CPU and a GeForce RTX 3090 GPU.

\subsection{Graph-Graph Classication Task}
For the graph-graph classification tasks, we conducted five experiments and reported average and standard deviation of AUC values in Table 2. Our Transformer-based Graph Fusion Model (GFM) excelled on all datasets. For example, on the OpenSSL dataset, GFM outperformed the H2MN(RW) by 1.34\%, 0.76\%, and 2.2\%, nearing 100\% accuracy. This showcases the superiority of our model's graph fusion and multi-view similarity over H2MN's subgraph matching, emphasizing the importance of detailed interaction and diverse correlation levels in graph similarity learning.

Additionally, substituting the Transformer with a Performer, which maintains linear computational complexity, still outperformed benchmarks. This success is attributed not only to the Transformer's capabilities but also to our graph fusion algorithm's efficiency in processing graph interaction information. This method, suitable for parallel encoding, enhances both global and local similarity learning, and is extendable to recommendation systems for graph-level interaction encoding.

\subsection{Graph-Graph Regression Task}

In the graph-graph regression tasks, we repeated five experiments and summarized their average performance in Table 3 using three metrics:\textit{\textbf{ Mean Squared Error (MSE)}}, \textit{\textbf{Spearman's rank correlation coefficient ($\mathbf{\rho}$)}}, and \textit{\textbf{precision@10 (p@10)}}. These metrics assess the squared differences, ranking correlation, and top 10 matches between predicted and true values, respectively.

Results mirror those from the classification tasks, with our Transformer and Performer models performing comparably to advanced baselines and excelling on the LINUX dataset as detailed in Table 4. The superior performance of these GNN methods over approximate algorithms like VJ, Beam, and HED underscores the GNN framework's effectiveness.

\subsection{Efficiency}

\begin{figure}
	\centering
	\includegraphics[width=0.99\linewidth]{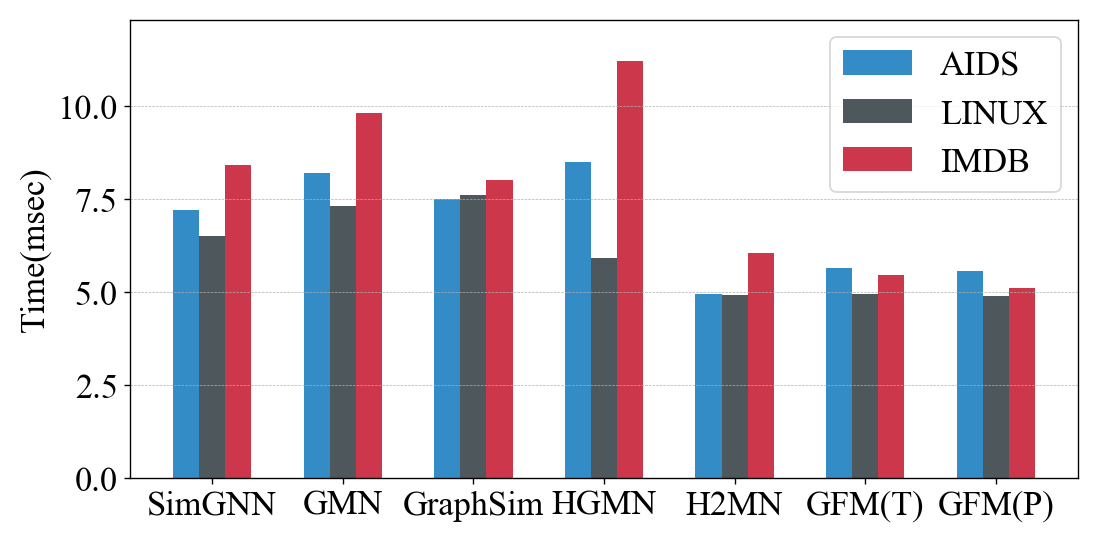}
	\caption{Running time comparisons (average time consumption on
		one pair of graphs in milliseconds)}
	\label{fig:screenshot006}
\end{figure}

To highlight the efficiency of our model, we analyzed the runtime of various methods in Figure 3. As shown, all models had their hidden layer nodes set to 48. Among them, the HGMN model performed the worst on large graph datasets, likely due to the time-consuming recursive computation structure of its BiLSTM. The H2MN model, which incorporates a pooling algorithm, enhanced computational efficiency compared to other models. The model proposed in this paper, which simply consists of a Transformer and a one-dimensional convolution, has a straightforward structure and shows improved computational efficiency over previously proposed algorithms.

\subsection{Complexity Analysis}
In this subsection, we examine the computational complexity of our model. Node embeddings are learned using the GraphSAGE layer, with time complexity \(\mathcal{O}(\max(|\mathcal{E}_{i}| + |\mathcal{E}_{j}|))\), where \(|\mathcal{E}1|\) and \(|\mathcal{E}2|\) represent the edge counts in the respective graphs. The Transformer module in the graph fusion phase has time complexity \(\mathcal{O}(|\mathcal{V}|^2 f)\) and space complexity \(\mathcal{O}(|\mathcal{V}|^2 + |\mathcal{V}| f)\), with \(|\mathcal{V}|\) and \(f\) denoting node count and feature dimension, respectively. The Performer reduces space complexity to \(\mathcal{O}(|\mathcal{V}|r + |\mathcal{V}|f + rf)\) and time complexity to \(\mathcal{O}(|\mathcal{V}|rf)\), where \(r\) is a Performer-specific parameter, generally less than \(d\). Node similarity calculation has complexity \(\mathcal{O}(|\mathcal{V}|f)\). These complexities affirm the high computational efficiency of our proposed model.

\begin{table}[h]
	\centering
	\caption{Results of ablation experiment. We use bold to highlight wins (mse is in the scale of $10^{-3}$).}
	\label{tab:my_label}
	\begin{tabular}{@{}l|cc|cc|cc@{}}
		\toprule
		\textbf{Datasets} & \multicolumn{2}{c}{\textbf{AIDS}} & \multicolumn{2}{c}{\textbf{LINUX}} & \multicolumn{2}{c}{\textbf{IMDB}} \\ 
		\cmidrule(r){2-3} \cmidrule(lr){4-5} \cmidrule(l){6-7}
		\textbf{Evaluation} & {\textbf{mse}} & {\textbf{p@10}} & {\textbf{mse}} & {\textbf{p@10}} & {\textbf{mse}} & {\textbf{p@10}} \\
		\midrule
		w/o Graph-sim         & 1.423          & 0.616           & 0.142         & 0.989          & 0.627         & 0.854          \\
		w/o Node-sim         & 1.249 & 0.634          & 0.333 & 0.970           & 0.684          & 0.839           \\
		
		w/o Graph Fusion        & 2.135 & 0.499  & 0.138 & 0.988  & 0.531 & 0.869           \\
		\textbf{GFM(Transformer)}               & \textbf{1.114} & \textbf{0.671}  & \textbf{0.058} & \textbf{0.999}  & \textbf{0.321} & \textbf{0.879}  \\
		\bottomrule
	\end{tabular}
\end{table}

\subsection{Ablation Study}
This paper introduces four innovative modules for assessing graph similarity in graph neural networks: graph fusion using the Transformer model, and three downstream modules for computing similarity vectors at graph-level, node-level, and cross-level between graphs and nodes. We conducted ablation studies by individually omitting each module, with results presented in Table 4 for three graph-graph regression datasets. These results highlight the distinct contributions of each module to the overall model performance. Specifically, the graph fusion module significantly influences performance on the AIDS dataset, whereas the node similarity calculation module has the most substantial effect on the LINUX dataset.

\subsection{Graph Search Case Study}

\begin{figure}
	\centering
	\includegraphics[width=0.99\linewidth]{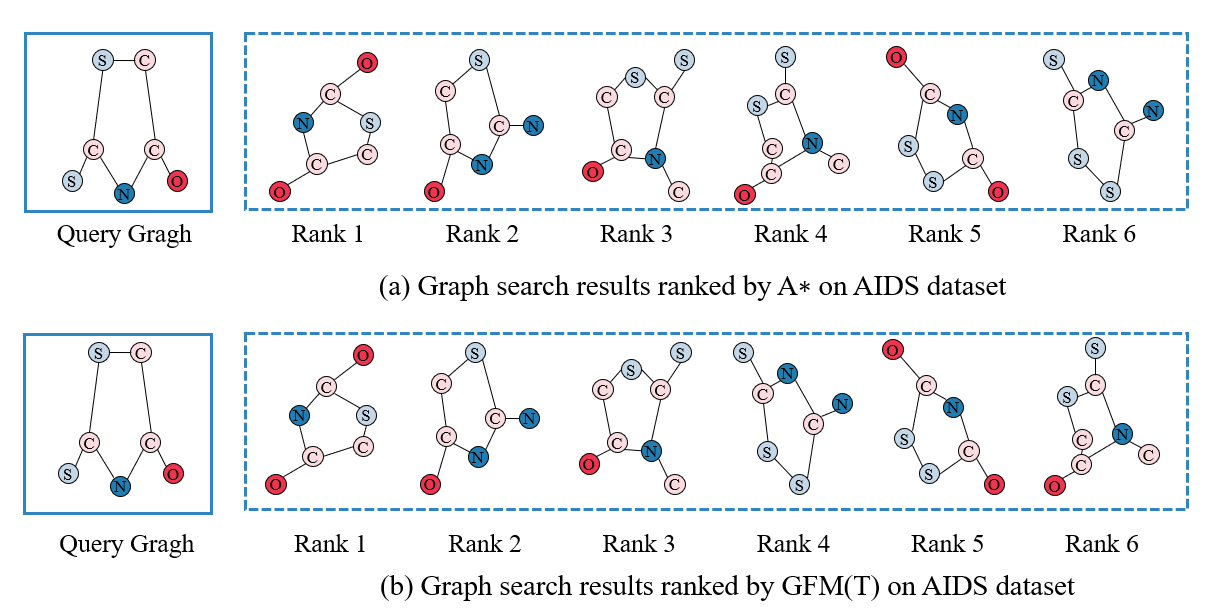}
	\caption{Graph search case study on AIDS dataset}
	\label{fig:screenshot004}
\end{figure}

In this section, we demonstrate the application of our model in graph retrieval tasks using graph-graph regression results. The task involves retrieving the top 6 most similar graphs from a dataset, given a query graph. An example from the AIDS dataset is depicted in Figure 4, where (a) presents the query graph alongside the top 6 ground truth results as determined by the A* algorithm, and (b) displays the retrieval results of our model. The close alignment of our model's results with the A* algorithm's ground truth rankings underscores our model's practical effectiveness.

\section{Conclutions}
In this paper, we addressed the challenge of learning graph similarity in graph-graph classification and regression tasks, necessitating joint reasoning about input graph pairs. We introduced a novel graph fusion algorithm that enables simultaneous encoding of graph pairs with linear complexity. This algorithm merges two graphs into one large graph, utilizing a global attention mechanism for effective information fusion. Moreover, we developed two distinct modules for calculating graph similarity at graph-level and node-level between graph and node, providing a multi-perspective approach to similarity assessment. Our end-to-end method allows for the evaluation of any structured graph pair. Extensive testing on five popular datasets confirms our model's superiority over leading baselines.

%\newpage
% your own custom macro containing the $\backslash${\tt{includegraphics}} command to make things
%\bf{If you include a photo:}\vspace{-33pt}
%\begin{IEEEbiography}[{\includegraphics[width=1in,height=1.25in,clip,keepaspectratio]{fig1}}]{Michael Shell}
%Use $\backslash${\tt{begin\{IEEEbiography\}}} and then for the 1st argument use $\backslash${\tt{includegraphics}} to declare and link the author photo.
%Use the author name as the 3rd argument followed by the biography text.
%\end{IEEEbiography}

\vfill

\end{document}